\documentclass[12pt]{article}

\usepackage{epsfig}
\usepackage{a4}

\newcommand{\be}{\begin{equation}} 
\newcommand{\ee}{\end{equation}} 
\def\nuc#1#2{\mbox{$^{#1}${#2}}}
\newcommand{\ddxhe}{$d\,d\to\, $\nuc{4}{He}$\,X^0$}
\newcommand{\ddetahe}{$\vec{d}\,d\to\, $\nuc{4}{He}$\,\eta$}
\newcommand{\dpoldxhe}{$\vec{d}\,d\to\, $\nuc{4}{He}$\,X^0$}

\newcommand{\pdhex}{$p\,d\to\,$\nuc{3}{He}$\,X^0$}
\newcommand{\pdhx}{$p\,d\to\,$\nuc{3}{H}$\,X^+$}

\newcommand{\pndx}{$pn\to d\,X^0$}
\newcommand{\ppppx}{$pp\to pp\,X^0$}
\newcommand{\pndpi}{$pn\to d\pi^0$}
\begin{document}
\setlength{\unitlength}{1mm}
\mbox{ }
\vspace{20mm}

\includegraphics{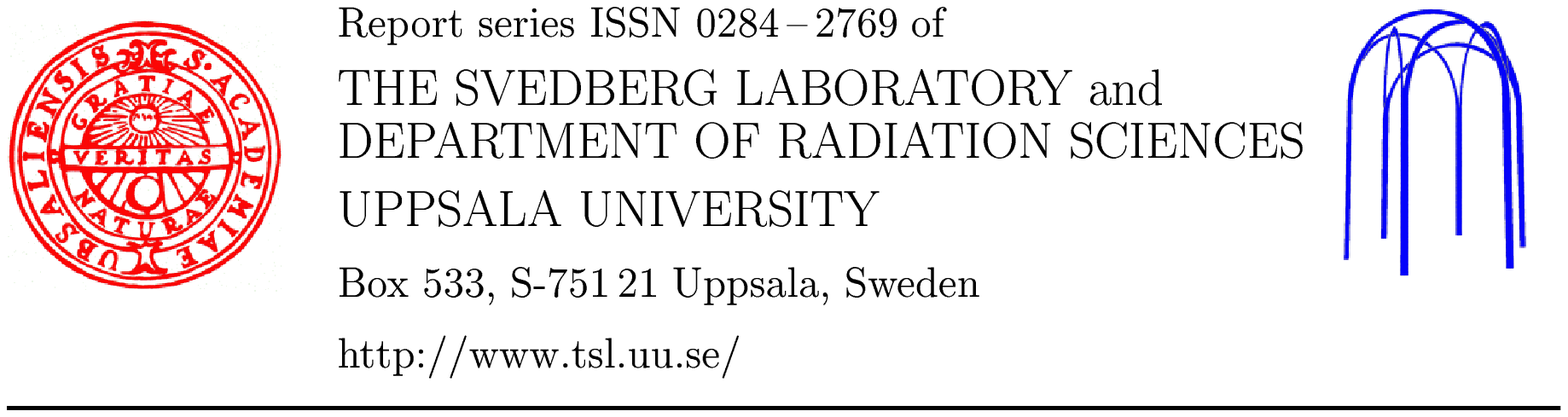}

\begin{flushright}
\begin{minipage}[t]{37mm}
{\bf TSL/ISV-98-0197 \\
October 1998}
\end{minipage}
\end{flushright}

\vspace{5mm}

\begin{center}
{\LARGE\bf Study of the ABC enhancement in the}\\[2ex] 
{\LARGE\bf \mbox{\boldmath $\vec{d}\,d \to \alpha\,X^0$} reaction}\\[4ex] 
R.~Wurzinger$^{1,2}$, 
O.~Bing$^3$, 
M.~Boivin$^2$, 
P.~Courtat$^1$, 
G.~F\"{a}ldt$^4$, 
R.~Gacougnolle$^1$, 
A.~G{\aa}rdestig$^4$\footnote{e-mail: grdstg@tsl.uu.se},
F.~Hibou$^3$, 
Y.~Le~Bornec$^1$, 
J.M.~Martin$^1$, 
F.~Plouin$^{2,5}$, 
B.~Tatischeff$^1$, 
C.~Wilkin$^6$\footnote{e-mail: cw@hep.ucl.ac.uk}, 
N.~Willis$^1$,
J.~Yonnet$^{1,2}$, 
A.~Zghiche$^3$\\[4ex]

$^1$ {\small\it Institut de Physique Nucl\'eaire, IN2P3/Universit\'e Paris-Sud,
 F--91406 Orsay Cedex, France}\\
$^2$ {\small\it Laboratoire National Saturne, F--91191 Gif-sur-Yvette 
 Cedex, France}\\
$^3$ {\small\it Institut de Recherches Subatomiques, IN2P3--CNRS / 
 Universit\'e Louis Pasteur, B.P.28, F--67037 Strasbourg 
 Cedex~2, France}\\
$^4$ {\small\it Nuclear Physics Division, Uppsala University, 
 Box 535, 751 21 Uppsala, Sweden}\\
$^5$ {\small\it LPNHE, Ecole Polytechnique, F-91128 Palaiseau, France}\\
$^6$ {\small\it University College London, London WC1E 6BT, U.K.}\\[6ex]
\end{center}

\vspace{3mm}

{\bf Abstract:}
The $\vec{d}\,d \to \alpha\,X^0$ reaction at beam energies close to the $\eta$
threshold shows very strong structure in the missing mass corresponding to the
ABC enhancement. The deuteron tensor analysing power $A_{yy}$, and the slope
of the vector analysing power $A_y$ with respect to angle, have been measured
for this reaction around the forward direction. Both signals are small, and
their variations with the $\alpha$-particle momentum are in broad agreement
with a theoretical model in which each pair of nucleons in the projectile and
target deuterons undergoes pion production through the $NN \rightarrow d\pi$
reaction. \\[2ex]

%\noindent
%Corresponding author:\\C.~Wilkin, Department of Physics \& Astronomy,\\
%University College London, Gower St., London WC1E 6BT, U.K.\\
%E-mail: cw@hep.ucl.ac.uk

\vfill
\clearpage
\setcounter{page}{1}

\newpage
\baselineskip 4ex

The discovery of a sharp enhancement in the missing mass spectrum of the
\pdhex\ reaction almost forty years ago~\cite{ABC} excited great interest since
it seemed to indicate either an enormous pion-pion scattering length (typically
3~fm~\cite{Spearman}) or a resonance in that system at only 30~MeV above 
the $2\pi$ threshold. The absence of any effect in the parallel \pdhx\ 
case showed that the anomalous behaviour was associated with isospin-zero 
pion pairs and hence presumably $s$-wave. However, the weight of evidence 
is that the isoscalar scattering length is small~\cite{PDG}, so that the 
ABC anomaly cannot be an intrinsic property of the two-pion system but must be
associated with the presence of other particles.

The characteristics of the ABC phenomenon were independently confirmed in the
\pdhex\ reaction~\cite{Banaigs1} and exactly the same type of effect was
observed in both \pndx~\cite{Banaigs2} and \ppppx\ at low $pp$ excitation
energies~\cite{Bergdolt}. The most splendid ABC manifestation is in the
\ddxhe\ reaction in the 1~GeV region, where it dominates a large fraction of
the differential cross section, with sharp peaks associated with its production
in the forward and backward c.m.\ hemispheres~\cite{Banaigs3,Nicole}. Since
the masses and widths of these peaks change somewhat with beam energy and
production angle, this reinforces the idea that the ABC might be some kind of
kinematic enhancement.

The simplest model for ABC production in \pndx\ involves the excitation of both
nucleons into $\Delta$-isobars through one pion exchange where, after the decay
of the two $\Delta$'s, the final neutron-proton pair sticks together to produce
the observed deuteron~\cite{Risser,Barry}. A theoretical description actually
looks simpler in the case of \ddxhe, where double-pion production can be
achieved by two pairs of beam and target nucleons undergoing a \pndpi\ reaction
(or similarly for charged pion production), as illustrated in
Fig.~1~\cite{anders}.

Each of the single-pion-production amplitudes can be driven by virtual $\Delta$
excitation, and this allows the large momentum transfer to be shared evenly
amongst all the nucleons. If the Fermi momenta in the initial deuterons are
neglected, then the $dd$ c.m.\ system is also that in the $np$ channel, which
means that the produced pions have the same energy. The dominantly $p$-wave 
nature of the \pndpi\ amplitudes, combined with the form factor coming from the
requirement that both deuterons stick to form an $\alpha$-particle, then leads
naturally to enhancements when the two pion momenta are parallel. These
correspond to the narrow ABC peaks with masses around 310~MeV and widths about
40~MeV. The model also predicts an enhancement when the pion momenta are
antiparallel, though this is somewhat suppressed by the much smaller ``sticking
factor'' when the two recoil deuterons are produced back to back~\cite{anders}.
The resultant broad central structure at the maximum missing mass of the 
reaction is a notable feature in all cases where the ABC is seen clearly
\cite{Banaigs1,Banaigs2,Banaigs3,Nicole}. By taking the dominant \pndpi\
amplitudes into account in the calculation, a good description of the \ddxhe\
spectrum and its angular dependence could be obtained~\cite{anders}. In order to
provide extra tests on this and other theoretical models, we report here upon an
experiment to measure the deuteron vector and tensor analysing powers in
\dpoldxhe\ in a small angular region around the forward direction.

The experiment was carried out at the SPESIII magnetic spectrometer at the 
Laboratoire National SATURNE (LNS) as a by-product of one to measure $\eta$
production in the \ddetahe\ reaction near threshold. Data were therefore
collected at several energies above the $\eta$-threshold at $T_d=1121$~MeV, but
also with one background run just below this energy. The experimental
conditions were identical to those described in ref.~\cite{Nicole} and so only
the essential details are reported here. The SATURNE accelerator delivered four
consecutive pulses of deuterons with different linear combinations of tensor
and vector polarisations quantised in the vertical ($y$) direction. These had
maximum values of $\rho_{20}=0.649\pm 0.011$ and $\rho_{10}=-0.405 \pm 0.011$
respectively~\cite{Egle}.

The wire chambers placed close to the focal plane of the spectrometer measured
the production angle of $\alpha$-particles in the horizontal ($x$) plane up to
$\theta_x=\pm 50$~mrad about the forward direction, but the only information on
the $y$-coordinate of the track was provided by collimators. In most of the runs
these subtended angles $\theta_y = \pm 20$~mrad and so the data were integrated
over this vertical angular domain. 

In Fig.~2 is shown a scatter plot of the momentum per unit charge $p/Z$ and
horizontal production angle $\theta_x$ of identified $\alpha$-particles 
arising from the interactions of unpolarised deuterons just below the $\eta$
threshold. The vertical bands close to the kinematic limits, and the less
intense one in the middle of the plot, are clear indications of the ABC peaks
and central bump respectively. Equally clear are the signs of the granularity
and inefficiencies in the detector system. It is important to note that these
are geometric in nature and rest unchanged for different polarisation states of
the incident beam. Summing this spectrum over $\theta_x$, the experimental
counting rate for the \ddxhe\ reaction is shown in Fig.~3a, from which it is
easier to identify the ABC peaks and central bump.

Although the vector analysing power $A_y$ has to vanish at $\theta_x=0$, the
experiment shows a clear left-right asymmetry over the horizontal opening
angle of $\pm 50$~mrad which is associated with the state of vector
polarisation of the beam. In the regions of the ABC peaks, where the statistics
are very good, it is possible to fit this directly with the form 
$A_y = c\,\theta_x$, allowing us to deduce the slope $c$ of the analysing
power near the forward direction. However, since the typical $A_y$ signal is 
$0.05$ or less, we attempt to compensate for this and the inefficiencies of 
the system by defining an average slope through
\begin{equation}
\overline{A_{y}'}= \frac{dA_y}{d\theta} = \int d\Omega\:
\left.\theta_x\: A_{y}\!\left(\frac{d^2\sigma}{d\Omega\,dp}\right)\right/
\int d\Omega\:(\theta_x)^{2}\left(\frac{d^2\sigma}{d\Omega\,dp}\right)\:\cdot
\end{equation}
The identification of this quantity with the slope of the analysing power at
$\theta=0$ is valid provided the angular integration domain is small, such that
$A_y$ varies linearly with $\theta_x$ and the differential cross section is
essentially constant at its value in the forward direction. To first order in
$(\theta_x, \theta_y)$ the form is not disturbed by the non-measurement of
$y$-component of the $\alpha$-particle production angle in SPESIII. This is a
reasonable assumption in the region of the ABC peaks, where the maximum
horizontal and vertical c.m.\ angles are about $17.5^{\circ}$ and $6^{\circ}$ 
respectively. However in the central bump, where the c.m.\ momenta are quite 
small, the integration in Eq.~(1) is over most of the available phase space and 
this has to be borne in mind in any subsequent interpretation.

By replacing the integrals in Eq.~(1) by sums over the experimental events,
using the polarised and non-polarised data respectively, the experimental value
of $\overline{A_{y}'}$ is deduced and shown in Fig.~3b. The data set taken just
below the $\eta$ threshold contained less than 10\% of the total available
counts and, in order to increase the statistical significance, results at six
different energies above and below the $\eta$ threshold ($1116.7 \leq T_d\leq
1127$~MeV) were summed in Fig.~3b. As a consequence the experimental points in
the middle of the central bump, $2000\leq p\leq 2100$~MeV/c, include some
$\eta$-production.

Only statistical errors in $\overline{A_{y}'}$ are shown in Fig.~3b. The
results depend weakly upon a possible off-set in the angular reading of 4~mrad
and, apart from a reduction in statistics, are not changed significantly by
cutting the range in $\theta_x$ from $\pm 50$~mrad to $\pm 30$~mrad. The
influence of potential systematic effects, arising from detector inefficiencies
or angular offset, could be tested by applying the same angular average as in
Eq.~(1) to evaluate the slope of the tensor analysing power $dA_{yy}/d\theta$,
which should vanish in the forward direction. Within statistical errors this
was found to be consistent with zero for all values of $p$ with a mean value of
$\langle\: \overline{A_{yy}'}\:\rangle = (-0.05\pm 0.10)$~rad$^{-1}$. The
overall systematic error on $\overline{A_{y}'}$ is hard to quantify but could be
rather larger than this due to the smallness of the signal.

In Fig.~3c are shown the values of the tensor analysing power $A_{yy}$
integrated over the whole horizontal and vertical aperture of SPESIII. Since a
slope determination is not required, the statistics from the single run at
$T_d=1116.7$~MeV, {\it i.e.}\ below the $\eta$ threshold, are here sufficient.
In addition to the statistical error bars, there is a systematic uncertainty of
only $\pm 2$\% arising from the polarisation of the beam~\cite{Egle}; it
should be noted that the value deduced for $A_{yy}$ for $\eta$ production in
this experiment is in agreement with the threshold theorem to this
accuracy~\cite{Nicole}. Detector inefficiencies in the ABC peaks could however
lead to slightly larger uncertainties. In this context it should be noted that
at strictly $0^{\circ}$ the values of $A_{yy}$ for the production of a $0^+$
state, such as the ABC, should be identical in the forward and backward peaks.
The difference in the peaks of Fig.~3c of $0.05\pm 0.02$ could in part arise
from integration over the finite SPESIII angular acceptance.

The analysing power signals are small and fluctuate with a similar frequency to
the counting rate shown in Fig.~3a. The shape of the latter is well reproduced
by the double-$\Delta$ model of ref.~\cite{anders} where the theoretical
calculation has been integrated over $|\theta_x|\leq 50$~mrad,  $|\theta_y|\leq
20$~mrad. The predictions in the central bump are the most uncertain since, in
this region, the $dd\to\alpha$ ``sticking'' factor is required for high relative
momenta. Furthermore, as noted in the top scale, the missing mass is well above
the three-pion threshold and extra pions could be produced. 

In the original double-$\Delta$ calculation of ref.~\cite{anders}, only the 
dominant $NN\to d\pi$ partial wave amplitude was considered and this gives  zero
for both the vector and tensor analysing powers. By including all the 
significant amplitudes of the C500 solution of ref.~\cite{Arndt}, the very 
encouraging estimates shown in Figs~3b, 3c at 1122~MeV could be achieved, 
where the results have been averaged over the experimental 
acceptance~\cite{anders2}. The refined model reproduces all the main features of
both $A_{yy}$ and $\overline{A_{y}'}$, the frequency of their fluctuations and
their strengths. The only gross discrepancy is an overall displacement of
$\overline{A_{y}'}$ by about 2~rad$^{-1}$, but this quantity is very sensitive
to any small extra  terms in the theoretical model. 

Our results on the vector and tensor analysing powers in \dpoldxhe\ give strong
quantitative support to the idea that $2\pi^0$ production in this reaction is
the result of single pion production occurring twice through the $pn\to d\pi^0$
reaction being repeated, or similarly for charged pions. Whether this remains
true closer to threshold, where the ABC peaks are not seen~\cite{Pia}, would
require the measurement of analysing powers over a greater range in energy. This
is no longer possible at Saturne following its final closure on December
2$^{nd}$ 1997.

We are grateful for all the help offered by colleagues at Saturne over many
years.

\newpage
\noindent
{\bf\large Figure Captions}\\[3ex]
{\bf Fig.~1 :} Double $NN\to d\pi$ model for the \ddxhe\ reaction\\[3ex]
{\bf Fig.~2 :}
Two-dimensional scatter plot of the momentum per unit charge $p/Z$ and
horizontal production angle $\theta_x$ of identified $\alpha$-particles 
arising from the unpolarised \ddxhe\ reaction just below the $\eta$ threshold at
$T_d=1116.7$~MeV.\\[3ex]
{\bf Fig.~3 :}\\
{\bf (a)} Raw counting rate for the unpolarised \ddxhe\ reaction at $T_d =
1116.7$~MeV, integrated over $\pm 20$~mrad in the vertical and $\pm 50$~mrad in
the horizontal directions, as a function of the $\alpha$-particle momentum. 
The curve is the unnormalised prediction of the double-$\Delta$ 
model~\protect\cite{anders,anders2}, integrated over the experimental 
acceptance. The top scale shows the values of the missing mass $M_X$.\\[1ex]
{\bf (b)} Average slope $\overline{A_{y}'}$ of the deuteron vector analysing
power of the \dpoldxhe\ reaction defined as in Eq.~(1). The experimental data
are an average over six beam energies with $1116.7 \leq T_d\leq 1127$~MeV,
whereas the solid curve is the prediction of the double-$\Delta$
model~\protect\cite{anders,anders2} at $T_d= 1122$~MeV. An arbitrary
displacement of the predictions by 2~rad$^{-1}$ gives the dashed curve, which
is a much better representation of the data.\\[1ex]
{\bf (c)} Deuteron tensor analysing power $A_{yy}$ of the \dpoldxhe\ reaction
at $T_d=1116.7$~MeV, averaged over the aperture of SPESIII. The theoretical
curve is the prediction of the double-$\Delta$ 
model~\protect\cite{anders,anders2}.
\newpage
\input epsf
\begin{figure}
\begin{center}
\mbox{\epsfxsize=4in \epsfbox{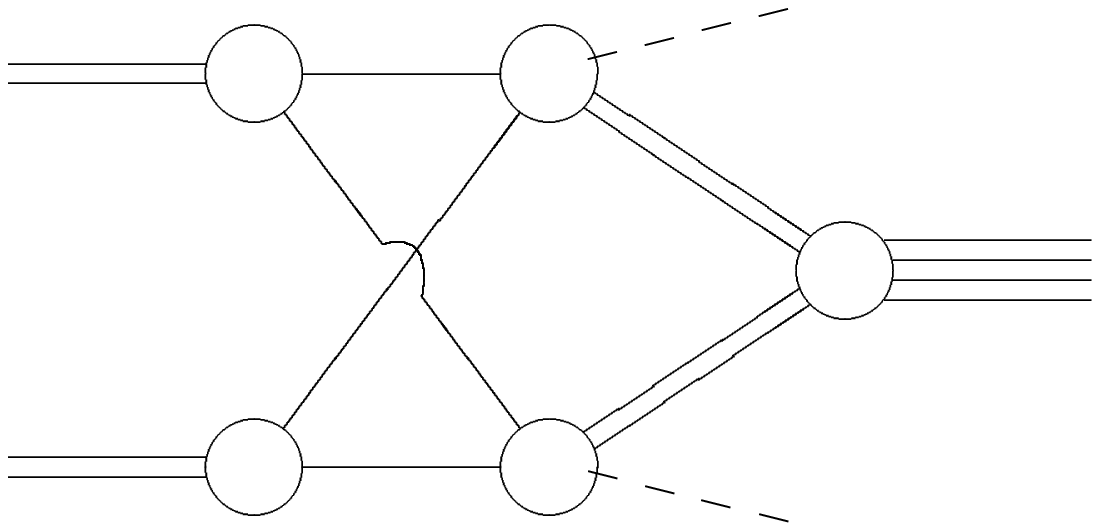}}
\end{center}
\caption{ }
\label{fig1}
\end{figure}

\newpage
\input epsf
\begin{figure}
\begin{center}
\mbox{\epsfxsize=5in \epsfbox{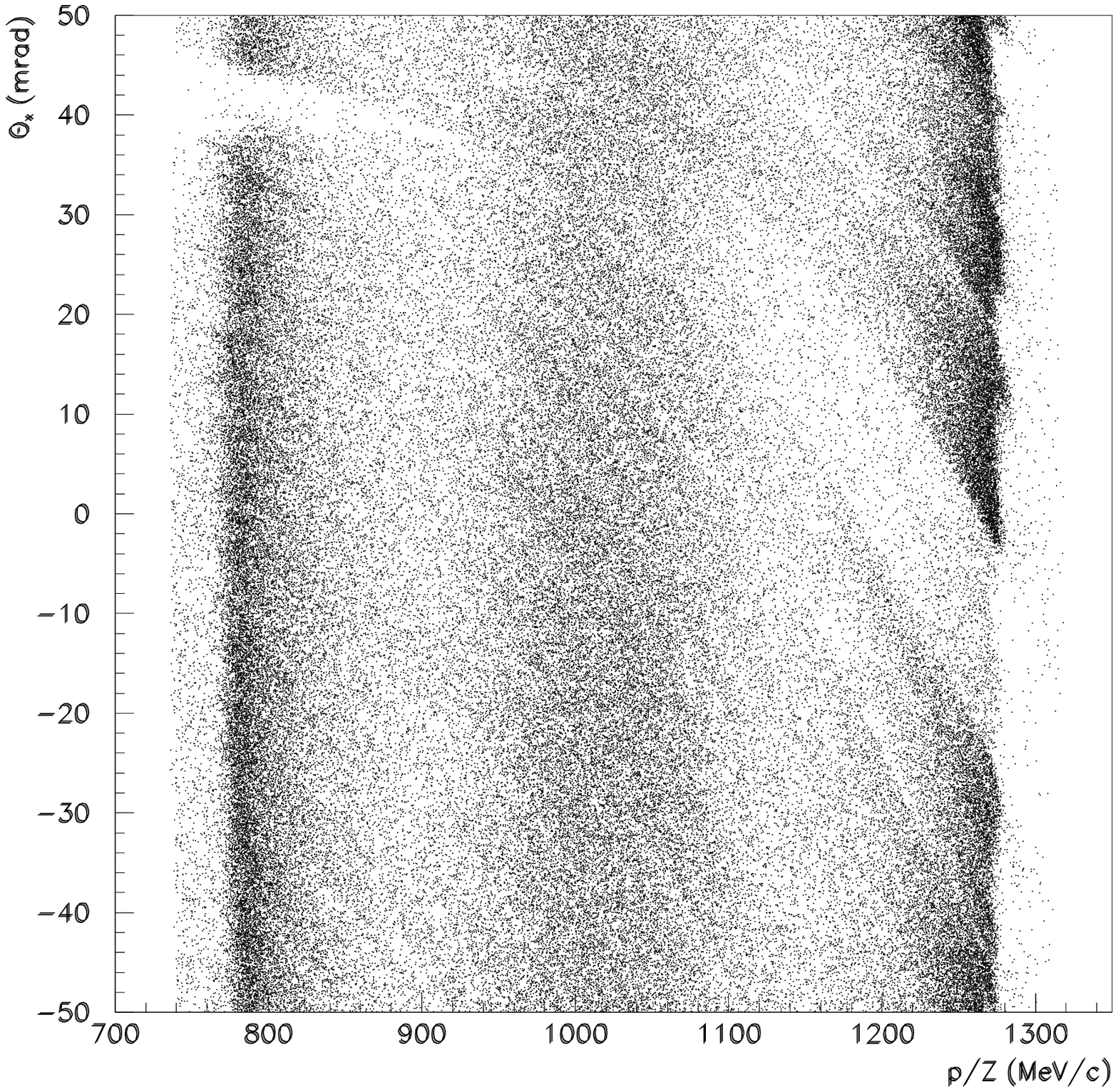}}
\end{center}
\caption{}
\label{fig2}
\end{figure}

\newpage
\input epsf
\begin{figure}
\begin{center}
\mbox{\epsfxsize=4.5in \epsfbox{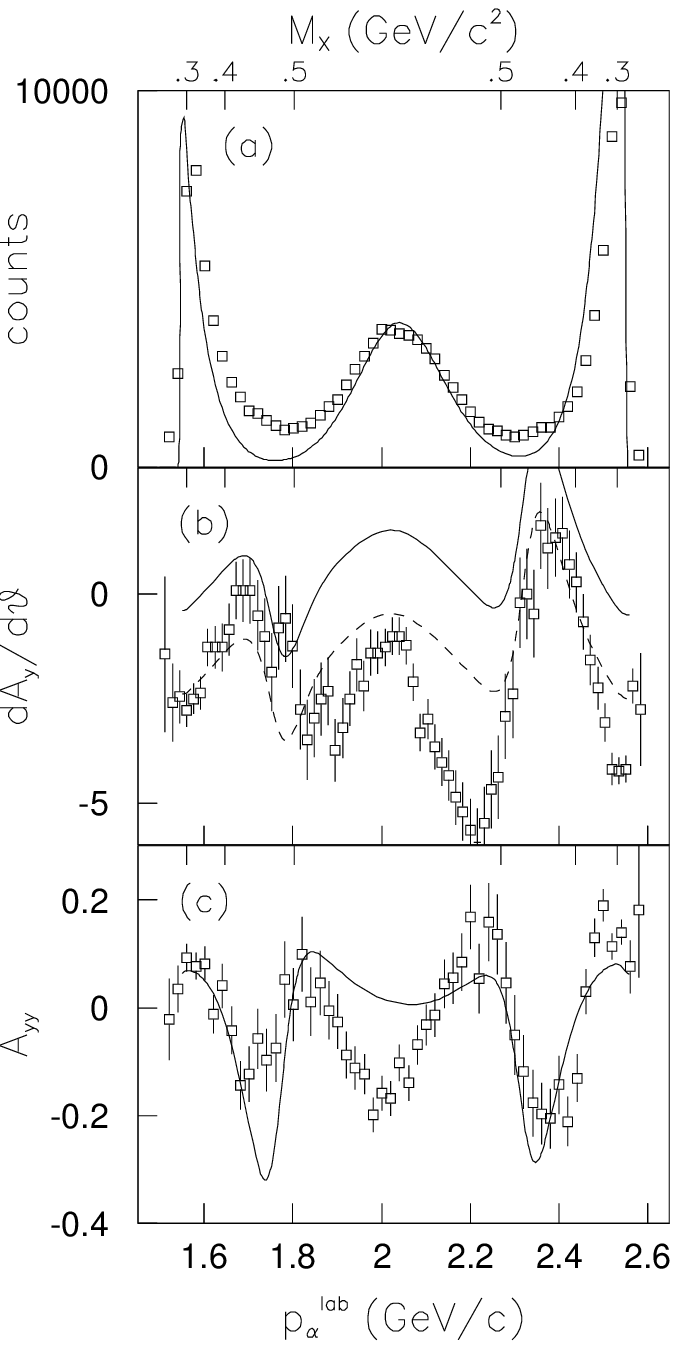}}
\end{center}
\caption{}
\label{fig3}
\end{figure}

\end{document}